\documentclass[epj]{svjour}
\usepackage{graphics}
\usepackage{hyperref}
\usepackage{xcolor}
\begin{document}
\title{Radial doses around energetic ion tracks and the onset of shock waves on the nanoscale}

\author{Pablo de Vera\inst{1,3,4}\thanks{Corresponding author: p.devera@qub.ac.uk}, Eugene Surdutovich\inst{2}, Nigel J. Mason\inst{3}, and Andrey V. Solov'yov\inst{4}\thanks{On leave from A. F. Ioffe Physical Technical Institute, 194021 St. Petersburg, Russian Federation}
}                     
\institute{School of Mathematics and Physics, Queen's University Belfast, BT7 1NN Belfast, Northern Ireland, United Kingdom \and Departament of Physics, Oakland University, Rochester, Michigan 48309, USA \and Department of Physical Sciences, The Open University, Walton Hall, MK7 6AA Milton Keynes, England, United Kingdom \and MBN Research Center, Altenh\"{o}ferallee 3, 60438 Frankfurt am Main, Germany}
\date{Received: date / Revised version: date}
%
\abstract{
Energetic ions lose their energy in tissue mainly by ionising its molecules. This produces secondary electrons which transport this energy radially away from the ion path. The ranges of most of these electrons do not exceed a few nanometres, therefore large energy densities (radial doses) are produced within a narrow region around the ion trajectory. Large energy density gradients correspond to large pressure gradients and this brings about shock waves propagating away from the ion path. Previous works have studied these waves by molecular dynamics simulations investigating their damaging effects on DNA molecules. However, these simulations where performed assuming that all energy lost by ions is deposited uniformly in thin cylinders around their path. In the present work the radial dose distributions, calculated by solving the diffusion equation for the low energy electrons and complemented with a semi-empirical inclusion of more energetic $\delta$-electrons, are used to set up initial conditions for the shock wave simulation. The effect of these energy distributions vs. stepwise energy distributions in tracks on the strength of shock waves induced by carbon ions both in the Bragg peak region and out of it is studied by molecular dynamics simulations.
}

\authorrunning{P. de Vera \textit{et al.}}

%
\maketitle
\section{Introduction}
\label{intro}

The interaction of energetic ion beams with biological materials is of great relevance in several disciplines including radiation protection and radiotherapy. The human body is bombarded by charged particles (i.e., protons, alpha particles) coming from natural radioactivity on Earth as well as during manned missions in space. Also, nuclear reactors produce artificial sources of energetic particles that can reach the operators. The ionising effects of these radiations can damage cells, therefore the dose delivered has to be monitored and unwanted exposure reduced by designing appropriate shielding~\cite{Cucinotta2006}. Nevertheless ionising radiation can also be used for therapeutic purposes and beams of energetic ions (mainly protons and carbon ions) have been used since the 1990s in the advanced radiotherapy technique known as ion-beam cancer therapy (IBCT) \cite{Schardt2010,Loeffler2013}. Ion beams feature an inverse depth-dose profile, where the energy loss reaches a maximum rate at low projectile velocities. This sharp maximum in the energy deposition is located at the very end of the ion trajectory and is known as the Bragg peak. The position of the Bragg peak for given ions and tissue depends on the initial energy of the ions. The latter can be tuned in such a way that the Bragg peak position overlaps with the tumour. This maximises the radiation impact on the tumour while sparing the surrounding healthy tissues. Therefore IBCT is an especially attractive technique for treating deeply seated tumours located near sensitive organs such as the brain stem or the optical nerve.

Apart from this main (macroscopic) feature, IBCT also has an increased cell-killing efficiency as compared to conventional radiotherapy (this is known as having an increased relative biological effectiveness) which arises from physico-chemical processes occurring on the nanoscale \cite{Solovyov2017}. Energetic ions lose their energy in tissue mainly by ionisation of its molecules. Most of the ejected secondary electrons have relatively low energies, indeed more than 80\% of them have energies below 50~eV~\cite{Surdutovich2014}. This only slightly depends on ion type, its energy and on the biological target~\cite{Surdutovich2009,deVera2013,deVera2013b}. The number of secondary electrons is roughly proportional to the linear energy transfer (LET) and tens of them can be produced on each nm of ion trajectory in the Bragg peak. Sub-45-eV electrons can travel only a few nanometers and therefore the energy lost by ions is deposited in a narrow region around their paths such that radial doses steeply decrease with the radii of the cylinder surrounding the ion path. 

The large number densities of reactive species induced by ionisation are prone to produce clustered patterns of damage in biomolecules surrounding the ion path. Such clustered damage in DNA molecules is likely to be lethal for cells and this increases the relative biological effectiveness of ion beams. Moreover, the energy deposition in narrow regions occurs very quickly, $\sim 50$ fs after ion passage, and cannot be effectively dissipated by other mechanisms, e.g. electron-phonon interaction or diffusion, ocurring on much longer times scales \cite{Surdutovich2015,Gerchikov2000}. Thus the rapid, intense  radial doses create the conditions for a strong explosion (very high pressure within a cylinder of $\sim$1-nm radius) that causes an onset of ion-induced shock waves~\cite{Surdutovich2010}. Such a dynamical response of the system on the nanometre scale changes {the accepted scenario with track structure followed by chemical stage \cite{deVera2017}.}

\begin{sloppypar}
These shock waves are an important part of the comprehensive scenario according to the predictions of the multiscale approach (MSA) to the physics of ion beam cancer therapy \cite{Surdutovich2014,Solovyov2009,Solovyov2017a}. The MSA is a phenomenon-based largely analytical method that is aimed at understanding the radiation damage with ions on a quantitative level. Its predictive power was recently demonstrated by the comparison of calculated cell survival probabilities for different doses, LET, oxygen concentrations and DNA repair efficiency levels with those obtained experimentally~\cite{Surdutovich2014,Verkhovtsev2016,Verkhovtsev2017}. The role of shock waves was included in these calculations.
One of the predicted effects of shock waves is a direct one, i.e., that ion-induced shock waves are capable of producing thermomechanical damage to biomolecules such as DNA if the latter are located sufficiently close to the ion path and LET is large enough~\cite{Surdutovich2013}. The other is related to the transport of reactive species produced in the vicinity of the ion path. According to the analysis of production and transport of reactive species (such as hydroxyl radicals and solvated electrons)~\cite{Surdutovich2015} in absence of shock waves these species cannot propagate fast enough to leave tracks and avoid reacting with each other. In contrast the collective flow of shock waves is instrumental in the swift transport of reactive species and is more effective (by a factor of about 80 for carbon ions around the Bragg peak~\cite{deVera2017b}) than diffusion. This spreads the radicals out to larger volumes increasing the radiation damage.

In previous work (\cite{Surdutovich2013,Yakubovich2011,Bottlander2015,deVera2016} and references therein) molecular dynamics simulations were used to explore the damaging effects of ion-induced shock waves on DNA molecules. They studied the energy deposited into covalent bonds by stress due to the shock wave at different values of LET. Covalent bonds located in the immediate vicinity to a track gained sufficient energy to be broken, corresponding to DNA strand breaks. This was observed in simulations for values of LET corresponding to Bragg peaks of heavy ions (heavier than Ar)~\cite{Surdutovich2013,Bottlander2015,deVera2016}. It was also shown~\cite{deVera2016,Yakubovich2011} that the wave front characteristics, obtained from molecular dynamics, perfectly agree with the features obtained from an analytical hydrodynamic model~\cite{Surdutovich2010}. However, thus far, the shock waves have not been observed experimentally.

All these simulations have been performed using the ``hot cylinder'' approximation in which the thermal energy that gives rise to shock waves is assumed to be uniformly distributed within a 1-nm-radius cylinder around the ion path. This radius was estimated as the average distance at which secondary electrons lose most of their energy, according to the random walk approximation to describe their motion~\cite{Surdutovich2013}.
Nevertheless the random walk approximation leading to the description of secondary electron transport by diffusion equations allows us to obtain the radial dose around the ion path. This permits us to set up the initial conditions for the simulations of the shock waves in a more realistic way {than the uniform energy distribution within a hot cylinder}.
\end{sloppypar}

In the present work {the diffusion equations are solved in order to describe the transport of sub-45-eV electrons and obtain the radial dose around the ion path, both in and out of the Bragg peak region. In addition to the sub-45-eV electrons more energetic electrons (referred to here as $\delta$-electrons) are included in the calculation of the radial dose by a spatially restricted LET equation~\cite{Xapsos1992}. This reproduces a large-radii tail of the radial dose characteristic for energetic ions. The resulting radial doses, at the end of the formation of track-structure but prior to shock wave development, are in good agreement with experimental data for tissue-equivalent gas and Monte Carlo simulations}. The addition of $\delta$-electrons eliminates the only adjustable parameter previously used for the calculation of the radial dose~\cite{Surdutovich2015}. These radial doses are then used to obtain the initial energy distributions for atoms in the molecular dynamics (MD) simulations of carbon ion-induced shock waves both in and out of the Bragg peak region.  The effect of these initial conditions is analysed and compared to previous simulations using the hot cylinder approximation. The strength of ion-induced shock waves and its dependence on LET is analysed.
The simulations were performed using the MBN Explorer simulation package \cite{Solovyov2012}.

The paper is organised as follows: section \ref{sec:Dr} describes the radial dose calculation based on solving the diffusion equation for sub-45-eV electrons and the inclusion of $\delta$-electrons contribution. The MD simulations of the ion-induced shock waves, using the radial doses obtained in section \ref{sec:Dr}, are described in section \ref{sec:MM}. The results of these simulations are given in section \ref{sec:results} where MD data are compared to the analytical hydrodynamic model. The final conclusions and remarks are given in section \ref{sec:conclusions}.

\section{Radial dose around energetic ion paths}
\label{sec:Dr}

The transport of sub-50-eV secondary electrons has been analysed in a number of works related to the multiscale approach (MSA) to the physics of radiation damage with ions~\cite{Surdutovich2014,Surdutovich2015,Solovyov2009,Solovyov2017a}. The random walk approximation leading to the diffusion equations used in these works adequately described their transport because their elastic as well as inelastic scattering cross sections were assumed to be isotropic in the first approximation. This approach has led to the calculation of the radial dose distribution \cite{Surdutovich2015}. The lack of attention to higher energy electrons has been justified by the fact that in all these works, the ions were considered to be in the Bragg peak region, where the fraction of higher energy electrons is small and the energy of these electrons is limited by kinematics. However, {out of the Bragg peak region, where the ions are more energetic,} a few very energetic $\delta$-electrons are capable of transferring a non negligible amount of energy far away from the ion path. Therefore a proper description of the radial dose {out of the Bragg peak region} must account for $\delta$-electrons.


In this work the vast majority of electrons (sub-45-eV electrons) are described by the latest implementation of diffusion equations for two generations of electrons (section~\ref{sec:RWA}). This more sophisticated approach is capable of describing the main effects produced by these numerous electrons in the MSA~\cite{Surdutovich2014,Surdutovich2015,Verkhovtsev2016}. In addition the $\delta$-electrons, although much less frequent, will be included here by a simpler methodology based on a spatially restricted LET formula \cite{Xapsos1992} (section \ref{sec:delta}). 
In section \ref{sec:LEEplusDelta} we explain how the contributions of sub-45-eV and $\delta$-electrons to the radial dose can be added and correctly weighted.

\subsection{The contribution of sub-45-eV electrons to the radial dose distribution}
\label{sec:RWA}

The transport of sub-45-eV electrons is described by diffusion equations that correspond to two generations of electrons \cite{Surdutovich2015}:
\begin{eqnarray}
\frac{\partial n_{1}(\vec{r},t)}{\partial t} & = & D_1\nabla^2 n_{1}(\vec{r},t)
 - \frac{n_{1}(\vec{r},t)}{\tau_1} \mbox{ , } \label{eq:diff1} \\
\frac{\partial n_{2}(\vec{r},t)}{\partial t} & = & D_2\nabla^2 n_{2}(\vec{r},t) + 2\frac{n_{1}(\vec{r},t)}{\tau_{1}}
 - \frac{n_{2}(\vec{r},t)}{\tau_2} \mbox{ , }
\label{eq:diff2}
\end{eqnarray}
where $n_{i}(\vec{r},t)$ are the electron densities at a location $\vec{r}$ (this vector connects the point of origin of the electron on the path and its observation point) and time $t$ for the first ($i=1$) and the second generation ($i=2$) of electrons. The more energetic electrons of the first generation (produced by the ion) undergo multiple elastic scattering (described by the first term on the r.h.s. of Eq. (\ref{eq:diff1})) before they lose their energy in inelastic collisions after an average time $\tau_1$ (second term on the r.h.s.) and thus form the second generation of electrons. The electrons of second generation also scatter elastically (first term on the r.h.s. of Eq. (\ref{eq:diff2})) until they finally lose most of their energy and thermalise or become low-energy electrons after an average time $\tau_2$ (third term on the r.h.s, of Eq. (\ref{eq:diff2})). These low-energy electrons are still capable of inducing DNA damage, but do not have sufficient energy to affect the radial dose distribution.  The electrons of the first and the second generations are assumed to have energies $W$ of $45$ and $15$~eV, respectively. The transition of an electron from the first to second generation is assumed to be an ionisation event in which an average energy $\overline{\omega} \approx 15$ eV is deposited to the medium and the remaining energy is equally distributed between the ionising and newly ejected electrons (second term on the r.h.s, of Eq. (\ref{eq:diff2})). The diffusion coefficients $D_i$ are related to the elastic mean free paths $\lambda_{{\rm el},i}$ by $D_i = v_i\lambda_{{\rm el},i} /6$, and the average times for inelastic collisions are related to the inelastic mean free paths $\lambda_{{\rm inel},i}$ by $\tau_i = \lambda_{{\rm inel},i} / v_i$, $v_i$ being the electron velocity of the corresponding generation. {We use the same values as in Ref.~\cite{Surdutovich2015}: $D_1 = 0.265$ nm$^2$/fs, $D_2 = 0.057$ nm$^2$/fs, $\tau_1 = 0.64$ fs and $\tau_2 = 15.3$ fs. The more energetic electrons have a longer elastic mean free path and a shorter inelastic mean free path, while the situation is reversed for the lower energy electrons.}

%
\begin{figure}
\resizebox{1.0\columnwidth}{!}{%
  \includegraphics{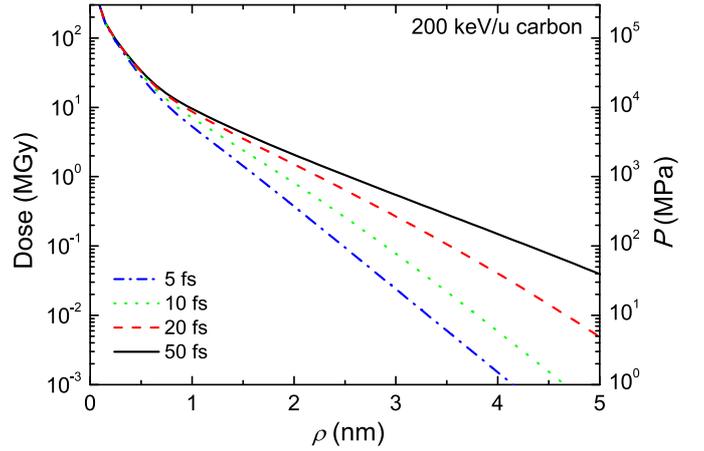}
}
\caption{(Color online) Radial dose produced in liquid water by a 200-keV/u carbon ion as a function of time after ion traversal. Lines show  present calculations using the diffusion equation for different times. The right axis represents the scale of pressure building up around the ion path.}
\label{fig:DrRWA}       
\end{figure}

The solutions of Eqs.(\ref{eq:diff1}) and (\ref{eq:diff2}) are given in detail in Ref.~\cite{Surdutovich2015}. They yield the electron densities for first and second generations as a function of time $t$ and the radial distance from the path $\rho$:
\begin{eqnarray}
n_1(\rho,t) & = & \frac{{\rm d}N(T)}{{\rm d}l} \frac{1}{4\pi D_1 t} \exp{\left( -\frac{\rho^2}{4 D_i t}-\frac{t}{\tau_1} \right)} \mbox{ ,} \label{eq:n1} \\
n_2(\rho,t) & = & 2\frac{1}{4\pi \tau_1} \frac{{\rm d}N(T)}{{\rm d}l} \int_0^{t} \frac{1}{D_1 t' + D_2(t-t')} \label{eq:n2} \\
 & \times & \exp{\left( -\frac{\rho^2}{4(D_1 t'+D_2(t-t')}-\frac{t-t'}{\tau_2}-\frac{t'}{\tau_1} \right)}{\rm d}t' \mbox{ , } \nonumber
\end{eqnarray}
where ${\rm d}N(T)/{\rm d}l$ is the number of electrons ejected per unit path length by an ion with kinetic energy $T$, i.e., the ionisation inverse mean free path \cite{deVera2013,deVera2013b}. From Eqs. (\ref{eq:n1}) and (\ref{eq:n2}) the radial energy deposition density profile at a given time $t$ is calculated by:
\begin{equation}
\frac{\partial \varepsilon(\rho,t)}{\partial t} = \overline{\omega} \left[ \frac{{\rm d}N(T)}{{\rm d}l} \delta^2(\rho) \delta(t) + \frac{n_1(\rho,t)}{\tau_1} + \frac{n_2(\rho,t)}{\tau_2} \right] \mbox{ . }
\label{eq:Dr}
\end{equation}
The second and third terms on the r.h.s. of Eq. (\ref{eq:Dr}) refer to the inelastic collisions of the electrons of the first and second generations, respectively, while the first term accounts for the energy deposited by direct ionisation by ion impact along the path, $\delta$ being delta functions in space and time. The energy deposition is simply related to the radial dose ${\cal D}(\rho,t)$ by ${\cal D}(\rho,t)=\varepsilon(\rho,t) / 2\pi \rho {\rm d}\rho l \varrho$, where $l$ is the ion path length and $\varrho$ is the mass density of liquid water.

%
\begin{figure}
\resizebox{0.853\columnwidth}{!}{%
  \includegraphics{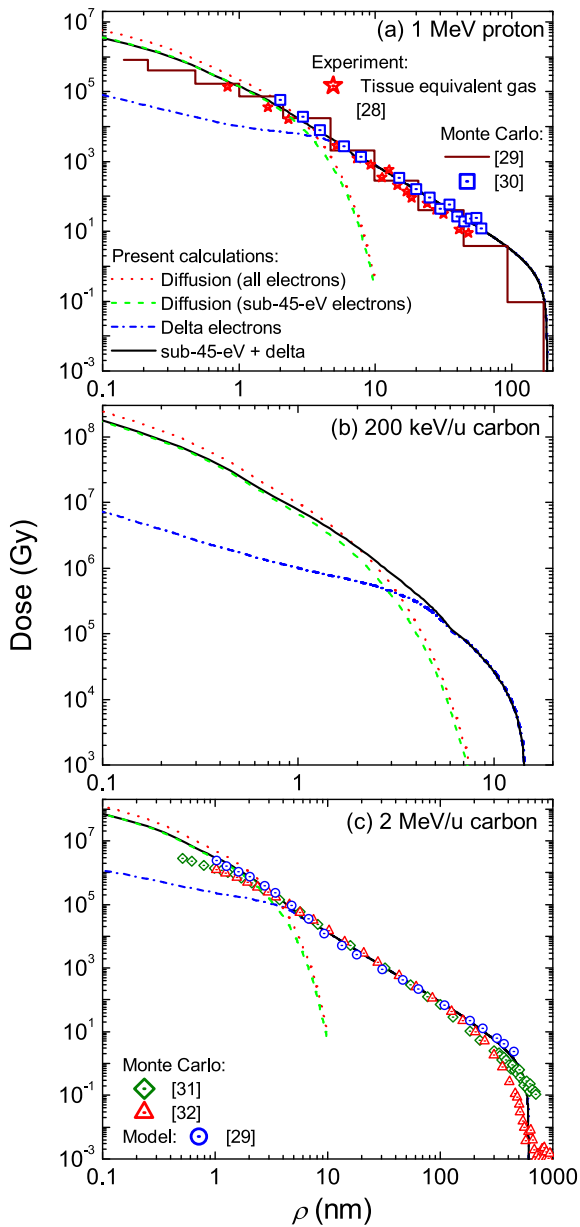}
}
\caption{(Color online) Radial doses produced in liquid water by (a) a 1-MeV proton, (b) a 200-keV/u carbon ion and (c) a 2-MeV/u carbon ion. Lines correspond to the present calculations (see the text for details), symbols are the results from Monte Carlo simulations \cite{Waligorski1986,Emfietzoglou2004,Liamsuwan2013,Incerti2014}, and stars depict experimental data for tissue-equivalent gas \cite{Wingate1976}.}
\label{fig:Dr}       
\end{figure}

%
\begin{figure}
\resizebox{0.8\columnwidth}{!}{%
  \includegraphics{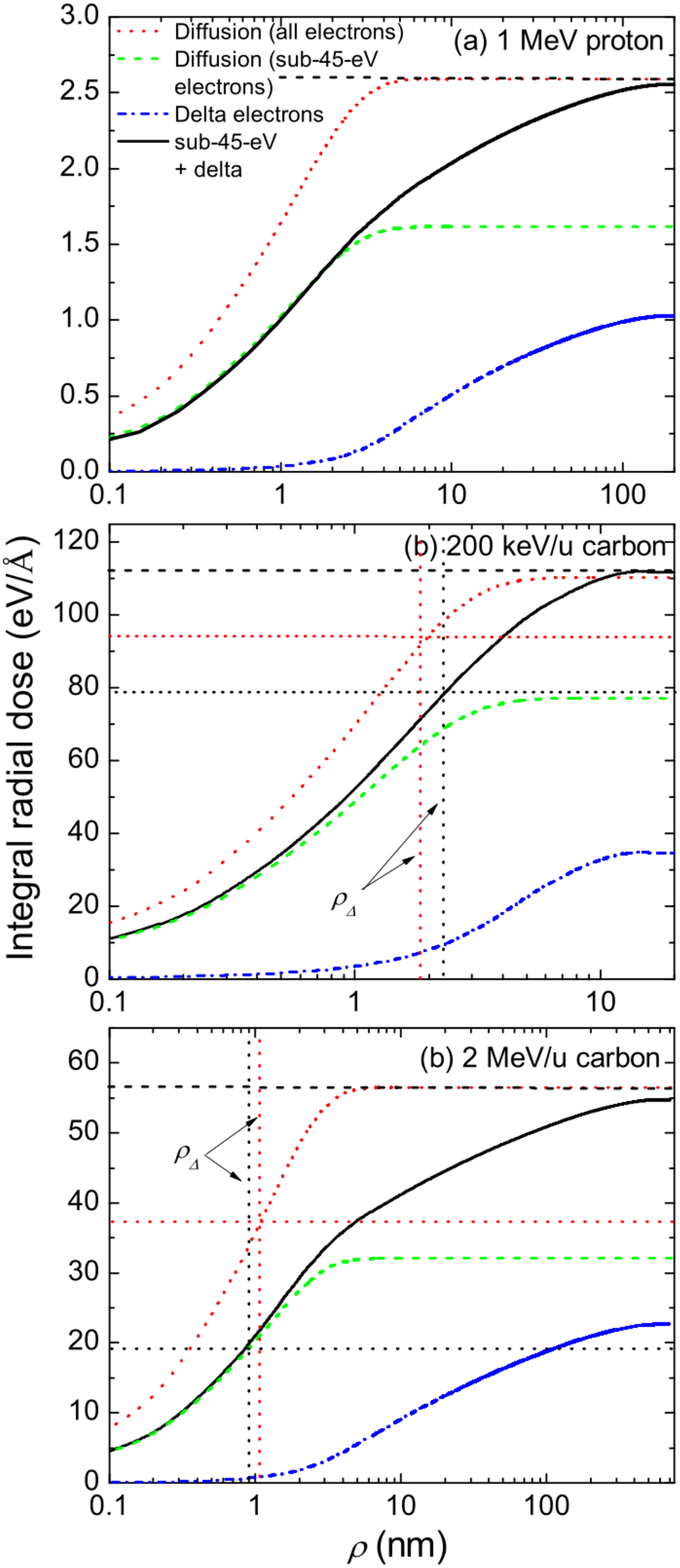}
}
\caption{(Color online) Integral of the radial energy deposited in liquid water by (a) a 1-MeV proton, (b) a 200-keV/u carbon ion and (c) a 2-MeV/u carbon ion. Vertical dotted lines mark the radius $\rho_{\Delta}$ corresponding to the restricted stopping power $S_{\Delta}$ as defined in section \ref{sec:results} (see the text for details).}
\label{fig:intDr}       
\end{figure}

Figure \ref{fig:DrRWA} shows the time evolution of the radial dose of a  carbon ion in the Bragg peak region ($T = 200$ keV/u). It takes approximately $\sim 50$ fs for all the electrons of the second generation to experience inelastic collisions, and then the radial dose converges. This time is short when compared to the characteristic times for the mechanisms capable of disipating these energies, such as electron-phonon interaction or diffusion, the latter being on the ps scale \cite{Surdutovich2015,Gerchikov2000}. The radial dose is built up much faster creating the conditions for a strong explosion of the medium on the ps scale.

As has been shown in Ref.~\cite{Surdutovich2015} the radial dose is equal to the pressure built up around the ion path. The pressure profile for a carbon ion in the Bragg peak region is represented in Fig. \ref{fig:DrRWA} with a scale on the right hand side axis. Such large pressures are enough to produce the hydrodynamic response of the liquid medium. The hydrodynamic treatment of the following expansion, which satisfies the conditions of ``strong explosion'', was given in Ref. \cite{Surdutovich2010}. The hydrodynamic treatment allows us to obtain useful physical characteristics of the ion-induced shock waves, and can serve as a benchmark for MD simulations~\cite{deVera2016}.

The volume integral of the radial dose energy per unit length along the ion path should converge to the total energy lost by the ion, in average, per unit path length, i.e., the electronic stopping power $S(T) = -\left<{\rm d}T(T) / {\rm d}l\right>$ \cite{Surdutovich2015}:
\begin{equation}
\int_0^{\infty} \int_0^{\infty} \frac{\partial \varepsilon(\rho, t)}{\partial t} 2\pi \rho {\rm d}\rho {\rm d}t = S(T) \mbox{ . }
\label{eq:intDr}
\end{equation}
{The diffusion equations are only appropriate to describe sub-45-eV electrons. As a consequence, if we try to describe the transport of all the electrons by diffusion (using their average energy $\overline{W} = 45$ eV for all of them) the average energy transferred per collision $\overline{\omega}$, which should be $\sim 15$ eV, has to be treated as an adjustable parameter that can be obtained from Eq.(\ref{eq:intDr}).}

This necessity to adjust $\overline{\omega}$ when $\delta$-electrons are not considered is exemplified by three different cases in Figs. \ref{fig:Dr} and \ref{fig:intDr}. Figure \ref{fig:Dr} shows, by dotted lines, the calculated radial doses in liquid water for (a) 1-MeV protons, (b) 200-keV/u carbon ions (energy in the Bragg peak region) and (c) 2-MeV/u carbon ions using the diffusion equation alone to describe all the electrons. The integral of the energy deposition as a function of the radius is also shown by dotted lines in Fig. \ref{fig:intDr}. The quantity $\overline{\omega}$ has to take different values in each case in order to integrate to the known stopping power \cite{GarciaMolinaPC}: this is 19 eV for 1-MeV protons, 15 eV for 200-keV/u carbon, and 20.5 eV for 2-MeV/u carbon. The corresponding stopping powers (obtained for ions in liquid water from the dielectric formalism \cite{GarciaMolinaPC,GarciaMolina2009,Abril2011,deVera2017c}) and $\overline{\omega}$-values for each case are summarised in Table \ref{tab:systems}. {It is worth noting that for carbon ions in the Bragg peak, $\overline{\omega} = 15$ eV, as expected. This is just a reflection of the fact that, in the Bragg peak, the role of $\delta$-electrons is not important. However, for more energetic ions, $\overline{\omega}$ deviates from 15 eV due to the fact that $\delta$-electrons are neglected.}  This free parameter will be removed when $\delta$-electrons are accounted for as will be described in the next section.

\begin{table}
\caption{Physical characteristics of ions of energy $T$, for which the radial doses have been calculated. The $\overline{\omega}$ values quoted are those needed for pure diffusion without consideration of $\delta$-electrons (this parameter always being 15 eV when including $\delta$-electrons, see the text for details). The stopping power $S$, the number of ejected electrons per nanometer ${\rm d}N /{\rm d}l$, as well as the number of sub-45-eV electrons ${\rm d}N_{45{\rm eV}} / {\rm d}l$, as used in the diffusion equations, are quoted.}
\label{tab:systems}       
\begin{tabular}{cccccc}
\hline\noalign{\smallskip}
Ion & $T$ & $S$ \cite{GarciaMolinaPC} & $\overline{\omega} \approx 15$ & ${\rm d}N / {\rm d}l$ & ${\rm d}N_{45{\rm eV}} / {\rm d}l$\\
 & (MeV/u) & (eV/\AA) & (eV) & (e$^-$/nm) & (e$^-$/nm) \\
\noalign{\smallskip}\hline\noalign{\smallskip}
H & 1.0 & 2.6   & 19.0 & 0.36 & 0.27 \\
C & 0.2 & 112.5 & 15.0 & 19.5 & 13.63 \\
C & 2.0 & 56.53 & 20.5 & 7.32 & 5.67 \\
\noalign{\smallskip}\hline
\end{tabular}
\end{table}

\subsection{Accounting for $\delta$-electrons}
\label{sec:delta}

The fact that $\overline{\omega}$ is an adjustable parameter in the diffusion equations described in the previous section is a consequence of normalisation of the radial dose to the stopping power, expressed by Eq. (\ref{eq:intDr}). Since in the diffusion description all the electrons are treated as if they had the average energies of 45 eV (first generation) or 15 eV (second generation), it cannot include the large radii tail of the radial dose arising from the energetic $\delta$-electrons. This tail is clearly seen for 1-MeV protons and 2-MeV/u carbon ions in Fig.~\ref{fig:Dr}, where experimental data for tissue-equivalent gas \cite{Wingate1976} and results from different Monte Carlo simulations and models~\cite{Waligorski1986,Emfietzoglou2004,Liamsuwan2013,Incerti2014} are depicted. In the following analysis we will account for the $\delta$-electron contribution by making use of a spatially restricted LET equation \cite{Xapsos1992}.



\begin{figure}
\resizebox{1.0\columnwidth}{!}{%
  \includegraphics{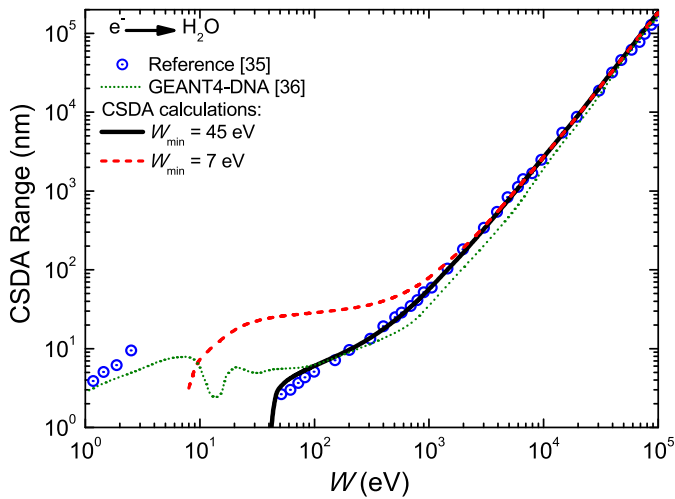}
}
\caption{(Color online) Continuous slowing down approximation (CSDA) range of electrons in liquid water as a function of their energy for different electron thermalisation energies $W_{\rm min}$ (solid and dashed lines). The dotted line represents GEANT4-DNA Monte Carlo results \cite{Francis2011} while symbols are a compilation of reference data \cite{Scifoni2010}.}
\label{fig:range}       
\end{figure}

In the first approximation the fraction of energy deposited within a microscopic cylinder, centered at the ion path, of radius $\rho$ by an ion of energy $T$ can be estimated within the LET (or restricted stopping power) approximation, given by~\cite{Xapsos1992,ICRU16}:
\begin{eqnarray}
f(T,\rho) & = & \frac{S_{\Delta}(T)}{S(T)} = \frac{\displaystyle{\int_0^{\Delta} (\hbar \omega) \frac{{\rm d}\Lambda(T)}{{\rm d}\omega} {\rm d}\omega}}{\displaystyle{\int_0^{W_{\rm max}} (\hbar \omega) \frac{{\rm d}\Lambda(T)}{{\rm d}\omega} {\rm d}\omega}} \nonumber \\ & \approx & \frac{\ln{\left(W_{\rm max}\Delta/I^2\right)}}{2\ln{(W_{\rm max}/I)}} {\rm ,}
\label{eq:f}
\end{eqnarray}
where the threshold energy $\Delta=\Delta(\rho)$ corresponds to the energy needed to produce an electron with range ${\cal R}=\rho$. ${\rm d}\Lambda(T)/{\rm d}\omega$ is the inelastic singly differential inverse mean free path as a function of the energy transfer $\hbar \omega$, $W_{\rm max}(T)=4\frac{m}{M}T$ is the maximum energy that is possible to transfer to a secondary electron of mass $m$ \cite{Rudd1992}, $M$ is the ion mass, and $I$ is the mean excitation energy of the material. $S_{\Delta}(T)$ denotes the restricted stopping power or LET \cite{ICRU16} while $S(T)$ is the stopping power of the ion of energy $T$. The last expression in Eq. (\ref{eq:f}) comes from using the Bethe formula for the stopping power \cite{Xapsos1992}.  The range of an electron with kinetic energy $W$ is usually calculated within the continuous slowing down approximation (CSDA) as:
\begin{equation}
{\cal R}(W) = \int_{W_{\rm min}}^{W}\frac{{\rm d}W'}{S(W')} \mbox{ , }
\label{eq:CSDAr}
\end{equation}
where $W_{\rm min}$ is the minimum thermalisation energy for electrons. The range-energy relation obtained by different models is given in Fig.~\ref{fig:range} for electrons in water. Symbols represent a compilation of reference data \cite{Scifoni2010}, while the dotted line are GEANT4-DNA Monte Carlo calculation results \cite{Francis2011}. The solid and dashed lines give the CSDA ranges calculated using Eq.(\ref{eq:CSDAr}) and the electronic stopping powers obtained from the dielectric formalism (as explained in Refs. \cite{deVera2011,GarciaMolina2016}, although in this case using the extended-Drude method, as explained, e.g., in Ref. \cite{deVera2013b}, accurate enough for high energies), employing a thermalisation energy of 45 and 7 eV, respectively. The decrease of the threshold energy produces an increase of the CSDA range. However, too low thresholds should be avoided in the current framework where the nuclear stopping power, important at low energies, is disregarded. The use of a threshold at 45 eV reproduces the high energy reference data quite well \cite{Scifoni2010} and is in good agreement with Monte Carlo results \cite{Francis2011}. This threshold is also useful for separating sub-45-eV and $\delta$-electron contributions as it is explained below, so we will use $W_{\rm min} = 45$ eV from now on.

%

In the LET approximation all the electrons with a range larger than the cylinder radius $\rho$ are assumed to escape it depositing no energy within it. This always underestimates the amount of energy deposited since it neglects (i) the energy deposited by electrons escaping from the cylinder on their way out and (ii) binding energies deposited by the electrons escaping from the cylinder. Xapsos \cite{Xapsos1992} suggested an extended spatially restricted LET equation in which these two contributions are accounted for by replacing $\Delta$ in Eq.(\ref{eq:f}) by $\Delta+\Delta_1+\Delta_2$. The parameters $\Delta_1$ and $\Delta_2$ phenomenologically increase the threshold energy due to reasons (i) and (ii) and virtually also increase the cylinder dimensions to account for the energy transfers missed by the LET approximation. These parameters were found by using simple arguments related to their expected asymptotic behaviour for small and large cylinders \cite{Xapsos1992}. The result is \cite{Xapsos1992}:
\begin{equation}
f_{\rm ion}(T,\rho) = \frac{\ln{\left(W_{\rm max}[\Delta+(1-\Delta/W_{\rm max}(\Delta+I)]/I^2\right)}}{2\ln{(W_{\rm max}/I)}} {\rm .}
\label{eq:XapsosLETf}
\end{equation}



It has been demonstrated that this approximation gives energy depositions within nanometric and micrometric volumes in accordance with Monte Carlo simulations \cite{Xapsos1992}. Using $f_{\rm ion}$, the dose deposited within a cylindrical shell of radius $\rho$ and width ${\rm d}\rho$ can be calculated as:
\begin{equation}
{\cal D}(T,\rho) = \frac{S(T)}{\varrho \pi l} \left[\frac{f_{\rm ion}(T,\rho+{\rm d}\rho)}{(\rho+{\rm d}\rho)^2} - \frac{f_{\rm ion}(T,\rho)}{\rho^2}\right] {\rm ,}
\label{eq:radial_dose}
\end{equation}
where $\varrho$ is the mass density of the target and $l$ the ion path length. 

\subsection{Separating sub-45-eV and $\delta$-electron contributions}
\label{sec:LEEplusDelta}

%
\begin{figure}
\resizebox{0.95\columnwidth}{!}{%
  \includegraphics{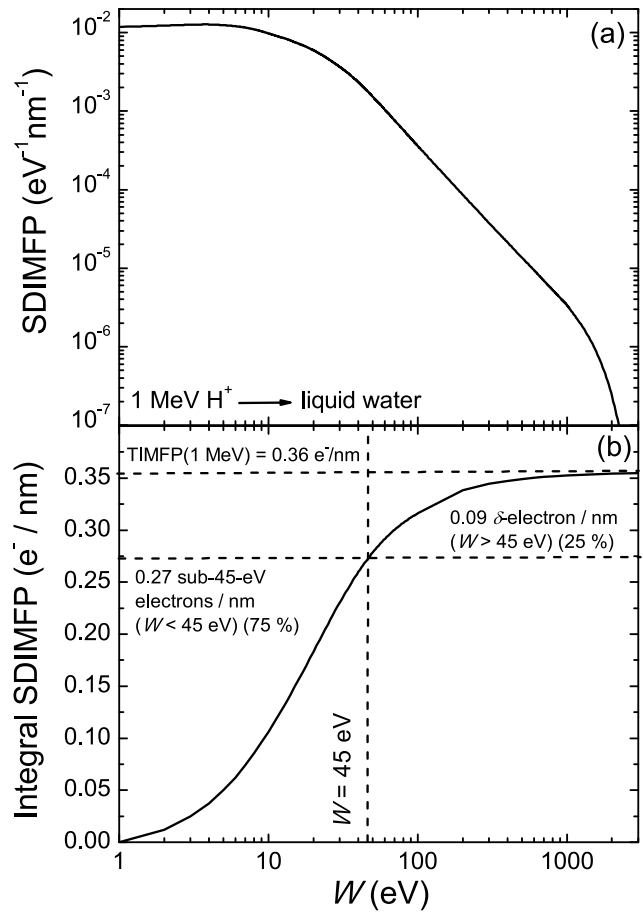}
}
\caption{(a) Ionisation singly differential inverse mean free path (SDIMFP) for 1-MeV protons in liquid water, and (b) integral of the SDIMFP as a function of the secondary electron energy $W$. Horizontal lines mark the total ionsation inverse mean free path (TIMFP) and the number of sub-45-eV electrons ejected per nanometer.}
\label{fig:SDCS}       
\end{figure}

It is tempting to simply sum the contributions from the diffusion equation for low radii (Eq. (\ref{eq:Dr})) and the spatially restricted LET equation for large radii (Eq. (\ref{eq:radial_dose})). However some care should be taken in order not to double count electrons in this description. The diffusion equation describes electrons below 45 eV, so we have to apply it only to the number of electrons ejected below this energy. Similarly, the spatially restricted LET equation must be forced to only describe electrons above 45 eV.

%
\begin{figure}
\resizebox{1.0\columnwidth}{!}{%
  \includegraphics{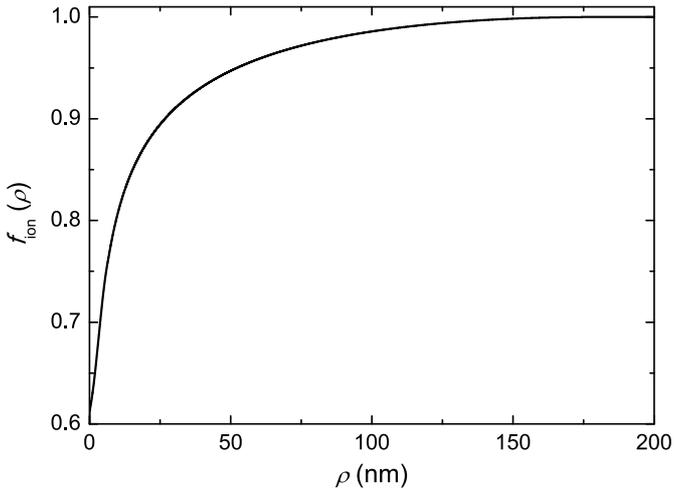}
}
\caption{Fraction of energy deposited by secondary electrons within a cylinder of radius $\rho$ around a 1 MeV proton path in liquid water.}
\label{fig:fion}       
\end{figure}

Regarding the sub-45-eV-electron contribution, it can be easily determined by using the singly differential ionisation inverse mean free path (SDIMFP), i.e., the energy spectrum of ejected electrons. This quantity can be calculated within the framework of the dielectric formalism \cite{deVera2013,deVera2013b}. A sample calculation is given for 1-MeV protons in liquid water in Fig. \ref{fig:SDCS}(a). The integral of the SDIMFP over energies up to $W$ gives the number of electrons ejected per unit path length with an energy equal or lower than $W$. This quantity is depicted for 1-MeV protons in water in Fig. \ref{fig:SDCS}(b), and it integrates to the total ionisation inverse mean free path (TIMFP). From this we can obtain the number of electrons ejected per unit path length with energies below 45 eV, ${\rm d}N_{45{\rm eV}}/{\rm d}l$, and use this quantity in the diffusion equations (\ref{eq:n1})--(\ref{eq:Dr}).

As can be seen in Fig. \ref{fig:SDCS}(b) out of the 0.36 electrons per nanometer ejected by 1-MeV protons, 0.27 (75 \%) of them have energy below 45 eV while the remaining 0.09 (25 \%) are ejected with energies above 45 eV. We will consider the former as sub-45-eV electrons which can be treated by the diffusion equations (using $\overline{\omega} = 15$ eV) while the latter will be regarded as high-energy electrons ($\delta$) and they will be treated by the spatially restricted LET formula.

The dashed lines in Fig.\ref{fig:Dr} show the radial doses calculated by the diffusion equations when they are only applied to the sub-45-eV electrons, as obtained from the calculated SDIMFP. For carbon ions in the Bragg peak, 19.5 electrons are ejected per nm, of which 13.63 are sub-45-eV (69.7 \%). For 2-MeV/u carbon ions, 5.67 (77.5 \%) electrons of the 7.32 electrons/nm ejected are sub-45-eV. All these quantities are summarised in Table \ref{tab:systems}. Figure \ref{fig:intDr} shows, by dashed lines, the integrals of these radial doses as a function of the radius. They are found to converge to values much lower than the stopping power, since the contribution from $\delta$ electrons has still to be included.

Regarding the $\delta$-electron contribution coming from the spatially restricted LET formula the calculation of the dose, Eq. (\ref{eq:radial_dose}), depends on the fraction of energy deposited $f_{\rm ion}$ given by Eq. (\ref{eq:XapsosLETf}). This expression in turn depends on the range-energy relation shown in Fig. \ref{fig:range}, which determines the cutoff energy $\Delta$ corresponding to each radius $\rho$. As discussed above the calculation of the CSDA range depends on the thermalisation energy $W_{\rm min}$ chosen for the secondary electrons. The solid line in Fig. \ref{fig:range} is calculated setting $W_{\rm min} = 45$ eV. This choice, apart from accurately reproducing the reference data for high energy electrons \cite{Scifoni2010}, also ensures every electron treated by the spatially restricted LET equation with $W\leq 45$ eV to have a range ${\cal R} \rightarrow 0$, as can be seen in Fig. \ref{fig:range}. The consequence of this is that all of these electrons will not be capable of moving from the ion path and they will deposit their energy exactly at $\rho=0$. This behaviour is seen in Fig. \ref{fig:fion}, where the fraction of energy $f_{\rm ion}$ deposited within a cylinder of radius $\rho$ is plotted. This fraction correctly converges to unity for long radii and goes to $\sim 0.6$ for $\rho=0$. This is the fraction of energy deposited at $\rho=0$ by all the electrons with energies $W\leq 45$ eV. Therefore, since we will treat these electrons by diffusion, the only thing we need to do is to disregard the energy deposited at $\rho=0$, as obtained with Eq. (\ref{eq:XapsosLETf}) when setting $W_{\rm min} = 45$ eV. In this way, the electrons with energies below 45 eV (sub-45-eV) are described by diffusion by virtue of the SDIMFP (dashed lines in Figs. \ref{fig:Dr} and \ref{fig:intDr}), while the spatially restricted LET equation is only describing the energy deposited by electrons above 45 eV ($\delta$), such that there is no double counting of electrons.

The $\delta$-electron contribution to the radial dose is shown by dash-dotted lines in Figs. \ref{fig:Dr} and \ref{fig:intDr}. As can be seen the $\delta$-electrons account for the large radii tail of the radial dose which compares fairly well with the experimental data for tissue equivalent gas for 1-MeV protons \cite{Wingate1976} as well as with the results of different Monte Carlo simulations and models for 1-MeV protons and 2-MeV/u carbon ions \cite{Waligorski1986,Emfietzoglou2004,Liamsuwan2013,Incerti2014}. The integral of the radial dose coming from $\delta$-electrons is shown in Fig. \ref{fig:intDr} by dash-dotted lines and accounts for an important fraction of the total energy deposited by the ion. This contribution is more modest for carbon in the Bragg peak, where it goes up to 30\% of the total deposited dose. However, the energy deposited by $\delta$-electrons is much more important for high energy ions going up to 40\% for 1-MeV protons and for 2-MeV/u carbon ions.

Finally, the sum of the sub-45-eV and $\delta$-electron contributions to the radial dose is given in Figs. \ref{fig:Dr} and \ref{fig:intDr} by solid lines. The resulting radial doses are in fairly good agreement with the experimental \cite{Wingate1976} and Monte Carlo data \cite{Waligorski1986,Emfietzoglou2004,Liamsuwan2013,Incerti2014} both for short and long radii. {Notice that these are the doses calculated at the end of the track-structure development and before the shock wave onset, so this is why they are in agreement with the cited experiments in the gas phase and Monte Carlo simulations, in which the shock wave does not appear.} The integral of the radial doses correctly converges to the stopping power without the need to adjust $\overline{\omega}$ anymore; this parameter is now fixed to 15 eV in all of the cases. All these facts are in favour of the chosen recipe to calculate radial doses accounting for the contributions of both sub-45-eV and $\delta$-electrons.

\section{Simulation of ion-induced shock waves}
\label{sec:MM}

In the previous section we have calculated radial doses at the end of the track-structure development. As has been mentioned this radial dose also corresponds to the pressure profile built up around the energetic ion path, large enough to prompt a hydrodynamic response of the liquid medium.
In previous works \cite{Surdutovich2013,deVera2016} shock waves were simulated using classical molecular dynamics (MD) but assuming that the radial dose distribution is a step function so that the energy is uniformly distributed inside a ``hot'' 1-nm-radius cylinder.
Here the obtained radial doses will be used to set up initial conditions for the MD simulations and the results will be compared with those obtained for hot cylinder based simulations, both in and out of the Bragg peak region. 

In MD simulations~\cite{Allen1989} the classical trajectories of all the atoms of the system, determined by their mutual interaction forces, are followed in time. The coordinates $\vec{r}_i(t)$ and velocities $\vec{v}_i(t)$ of each atom $i$ of mass $m_i$ are found at discrete time steps ${\rm d} t$ by numerically solving the Langevin equation:
\begin{equation}
m_i \frac{{\rm d}^2\vec{r}_i}{{\rm d}t^2} = \sum_{j\neq i} \vec{F}_{ij}-\frac{1}{\tau_{\rm d}}m_i \vec{v}_i+\vec{{\cal F}}_i \mbox{ , }
\label{eq:MD}
\end{equation}
in which $\sum_{j\neq i} \vec{F}_{ij}$ is the total force acting on atom $i$ due to its interaction with all the rest of atoms $j$ in the system (i.e., a system of coupled Newton's second law equations). The system of particles is kept close to a constant temperature ${\cal T}$ by coupling it to a Langevin thermostat through the second and third terms on the right hand side of Eq. (\ref{eq:MD}) which act as a viscous force on each particle of velocity $\vec{v}_i$ plus a random collision force. $\tau_{\rm d}$ is the damping time of the thermostat and $\vec{{\cal F}}_i$ is a random force with dispersion $\sigma_i^2=2m_i k_{\rm B}{\cal T}/\tau_{\rm d}$ with $k_{\rm B}$ being the Boltzmann's constant. The non pair-wise nature of the chemical bonding in biological molecules, where specific groups of atoms adopt given geometries, is accounted for in MD through the introduction of molecular mechanics potentials such as the CHARMM force field \cite{MacKerel1998}. In CHARMM, the force acting on the atom $i$ is obtained from the potential energy map $U(\vec{R})$ of the given set of coordinates $\vec{R}$ as $\sum_{j\neq i} \vec{F}_{ij} = \partial U(\vec{R})/\partial\vec{r}_i$ and comprises a combination of energies arising from the distances of bonds between pairs of atoms, the angles formed between groups of three sequentially bonded atoms, the dihedral and improper angles formed by groups of four bonded atoms as well as the nonbonding interactions between pairs of atoms, i.e., the pure Coulomb force and the van der Waals interaction.

In order to set up the MD simulations of shock waves, we built liquid water boxes of density 1~g/cm$^3$ of 17 nm distance from the center to the boundary ($x$--$z$ plane) and a length of $l = 4.346$ nm ($y$ direction) which were put into periodic boundary conditions~\cite{deVera2016}. The boxes were optimised and equilibrated at body temperature (${\cal T} = 310$ K) with MBN Explorer \cite{Solovyov2012} using the Langevin thermostat and the CHARMM force field, as explained in Ref.~\cite{deVera2016}.

The ion path is considered to cross the center of the box in the $y$ direction. The box may then be divided in concentric cylindrical shells of water molecules of 1~\AA \, thickness around the ion path. The velocities at equilibrium of atoms $i$ at each shell $j$ can be scaled by a parameter $\alpha_j$ depending on the amount of energy deposited in this shell as:
\begin{equation}
\sum_i^{N_j} \frac{1}{2}m_{i,j}(\alpha_j \cdot v_{i,j})^2 = \frac{3N_j k_{\rm B}{\cal T}}{2}+ f_j S l \mbox{ . }
\label{eq:Edepos}
\end{equation}
The first term on the right hand side of the equation corresponds to the initial kinetic energy of the atoms in the cylindrical shell $j$ (with $N_j$ atoms) at equilibrium (${\cal T} = 310$ K). The second term is the energy deposited by the ion in the shell $j$ when crossing the system which is its stopping power $S$ times the length of the simulation box $l$ times the fraction of the energy deposited in this shell $f_j$, as obtained from the radial dose.

Once the initial velocities of atoms of each cylindrical shell $j$ are determined according to Eq. (\ref{eq:Edepos}) the simulation of the shock wave is performed by turning off the thermostat in order to conserve the energy deposited by the ion in the medium. The results of these simulations are presented in the next section.


\section{Results and discussion}
\label{sec:results}

In this section we will analyse the results of MD simulations with initial conditions corresponding to the calculated radial dose distributions (see section \ref{sec:Dr}) compared to those with a step function energy distribution. Carbon-induced shock waves are discussed since carbon is one of the most employed ions in beam therapies and the shock wave effect is much more profound than in a proton-beam therapy.

Simulations have been performed for two energies: 200-keV/u, which is the energy in the Bragg peak region, and 2-MeV/u, which is a higher energy found out of the Bragg peak region. The radial doses for these two ions are shown in Fig. \ref{fig:Dr}(b) and (c). {The main effect of the radial dose is spreading the energy lost by the projectile out in the radial direction. This spreading is the most more important in the situation where more energetic ($\delta$) electrons are produced, i.e., for more energetic ions out of the Bragg peak region. There the strength of the induced shock waves is weakened compared to the hot cylinder approximation. To analyse these effects, three situations are analysed for each ion energy: (i) step-function radial energy distribution (hot cylinder approximation), (ii) radial dose-like distribution without $\delta$-electrons contribution, and (iii) radial dose-like distribution with $\delta$-electrons included. The different simulations performed are summarised in Table \ref{tab:hydrodynamics}. As discussed, the role of $\delta$-electrons is only expected to have influence for the 2-MeV/u case, since their contribution in the Bragg peak is small.}


\begin{table}
\caption{Summary of the MD simulations performed for carbon ion-induced shock waves. For 200-keV/u and 2-MeV/u carbon ions simulations have been done in the hot cylinder approximation as well as using the radial doses calculated without (pure diffusion) and with $\delta$-electrons. The restricted stopping powers $S_{\Delta}$ used for fitting the hydrodynamic equations and their ratio to the stopping power $S$ are quoted, as well as the radii corresponding to each restricted stopping power (as determined from Fig. \ref{fig:intDr}). See the text for details.}
\label{tab:hydrodynamics}       
\begin{tabular}{ccccc}
\hline\noalign{\smallskip}
$T$ & Simulation & $S_{\Delta}$ & $S_{\Delta}/S$ & $\rho_{\Delta}$ \\
(MeV/u) & type & (eV/\AA) &  & (nm) \\
\noalign{\smallskip}\hline\noalign{\smallskip}
0.2 & Hot cylinder & 112.5 & 1.0 & 1.0 \\
0.2 & Pure diffusion & 92.25 & 0.82 & 1.8 \\
0.2 & Diffusion$+\delta$ & 78.41  & 0.697 & 2.2 \\
2.0 & Hot cylinder & 56.53 & 1.0 & 1.0 \\
2.0 & Pure diffusion & 20.5 & 0.655 & 1.0 \\
2.0 & Diffusion$+\delta$ & 19.16 & 0.339 & 0.9 \\
\noalign{\smallskip}\hline
\end{tabular}
\end{table}

The strength of pressure waves generated in these three situations is analysed in comparison with the analytical hydrodynamic model developed in Ref. \cite{Surdutovich2010}. This approach, in which the energy is assumed to be deposited in an infinitely thin volume, predicts several physical properties of the shock waves such as the position of the wave front as a function of time:
\begin{equation}
R(t) = \beta \sqrt{t} \left[\frac{S}{\varrho}\right]^{1/4} \mbox{ , }
\label{eq:R_t}
\end{equation}
or the pressure on the wave front as a function of its radius,
\begin{equation}
P(R) = \frac{1}{2(\gamma+1)}\frac{\beta^4 S}{R^2} {\mbox .}
\label{eq:P_R}
\end{equation}
In both equations, $\varrho = 1$ g/cm$^3$ is the density of unperturbed liquid water, $\gamma=C_{\rm p}/C_{\rm v} = 1.222$, $S$ is the stopping power of the crossing ion and $\beta$ is a parameter whose value for liquid water is $\beta = 0.86$, as shown in Ref. \cite{Surdutovich2010}.

As it was demonstrated in Ref. \cite{deVera2016}, the properties of the wave front can be obtained from MD simulations. The pressure profile as a function of the radius, initially very sharp, propagates to longer radius as a function of time, gradually becoming weaker and less sharp. The maximum of these distributions can be regarded as the wave front of the pressure wave and its position and intensity can be directly compared to the results yielded by Eqs. (\ref{eq:R_t}) and (\ref{eq:P_R}). This allows the comparison of simulations.

\begin{figure}
\resizebox{0.9\columnwidth}{!}{%
  \includegraphics{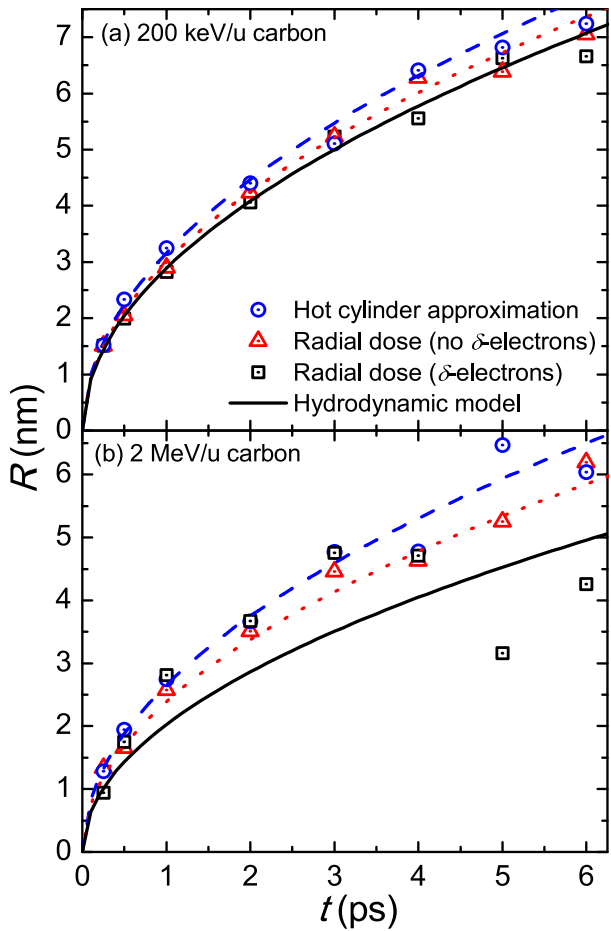}
}
\caption{(Color online) Position of the wave front as a function of time for (a) 200-keV/u and (b) 2-MeV/u carbon ions. Symbols are results from MD simulations, while lines correspond to the hydrodynamic model.}
\label{fig:Rt}       
\end{figure}

\begin{figure}
\resizebox{0.9\columnwidth}{!}{%
  \includegraphics{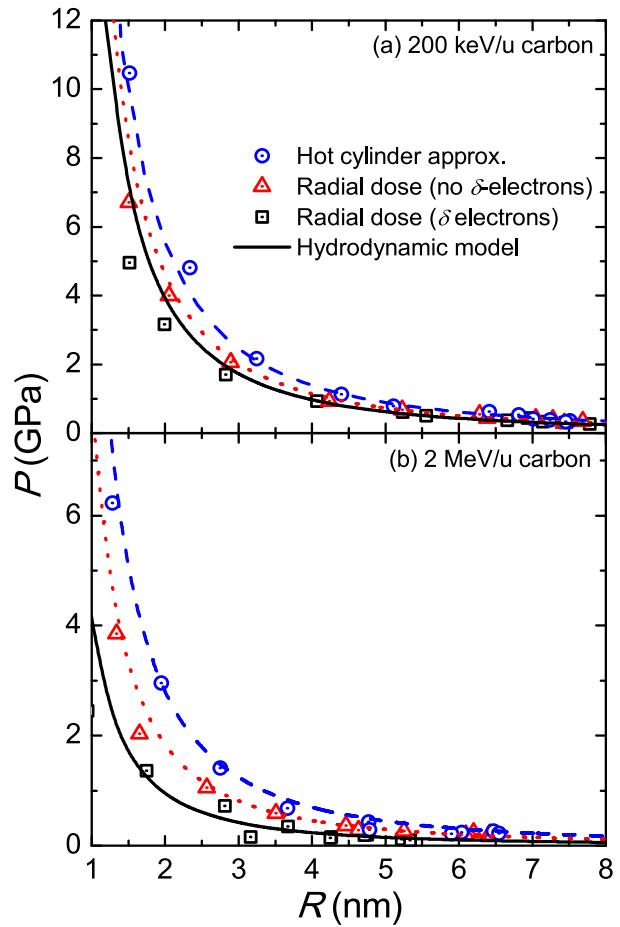}
}
\caption{(Color online) Pressure on the wave front as a function of its radial position for (a) 200-keV/u and (b) 2-MeV/u carbon ions. Symbols are results from MD simulations, while lines correspond to the hydrodynamic model.}
\label{fig:PR}       
\end{figure}

Figures \ref{fig:Rt} and \ref{fig:PR} show by symbols, respectively, the position and pressure on the wave front as obtained from MD simulations for (a) carbon ions in the Bragg peak region (200-keV/u) and (b) carbon ions out of the Bragg peak region (2-MeV/u). Circles show the results for simulations in the hot cylinder approximation. As was shown in Ref. \cite{deVera2016} the position and pressure on the front are reproduced very well by the analytical results. This is also seen in these figures where the dashed lines show the results of Eqs. (\ref{eq:R_t}) and (\ref{eq:P_R}) when the corresponding stopping powers of each ion are used, see Tables \ref{tab:systems} and \ref{tab:hydrodynamics}.

The results with initial energy distributions corresponding to the calculated radial doses are different. Let us start the discussion with the results obtained in the Bragg peak region (200-keV/u). Figs.~\ref{fig:Rt}(a) and \ref{fig:PR}(a) show, by triangles, the wave front properties when the radial dose obtained from pure diffusion is used (i.e., no $\delta$-electrons included); the squares indicate the results corresponding to the radial dose distributions including $\delta$-electrons. It can be seen that the speed of propagation of the wave front and its pressure are somewhat lower in these two cases as compared to the hot cylinder approximation. The intensity is only slightly smaller in the case of the radial dose with $\delta$-electrons included as compared with that without them.

These results, although lower in absolute value compared to the hot cylinder approximation simulations, seem to follow the same functional form. Thus, the hydrodynamic model is still useful to describe them. Nevertheless, owing to the lower intensity of these shock waves, the stopping power $S$ in Eqs. (\ref{eq:R_t}) and (\ref{eq:P_R}) has to be replaced by an effective stopping power which we will denote as $S_{\Delta}$. When $\delta$-electrons are not included (pure diffusion), $S_{\Delta}/S = 0.82$. When they are, $S_{\Delta}/S = 0.7$. Therefore, in the Bragg peak region the consideration of an accurate radial dose reduces the intensity of the shock wave, for a given position $R$, by up to 30\% compared to the hot cylinder approximation and the role of $\delta$-electrons is small as expected.

The situation is more drastic out of the Bragg peak region. As can be clearly seen in Figs. \ref{fig:Rt}(b) and \ref{fig:PR}(b), for 2-MeV/u carbon ions both the wave front velocity and pressure are heavily reduced when compared to the hot cylinder situation, when the radial dose is used, especially when the $\delta$-electrons are included. Without $\delta$-electrons $S_{\Delta}/S = 0.66$ but when they are included $S_{\Delta}/S = 0.34$. The pressures in this case are much lower than those in the Bragg peak case because of a smaller stopping power of 2-MeV/u carbon ions compared to 200-keV/u ions. Moreover the consideration of an accurate radial dose where $\delta$-electrons carry the energy further away reduces the intensity of shock waves, for a given position $R$, by $\sim$65\% compared to the hot cylinder approximation.

%

%

The obtained values for $S_{\Delta}$ (summarised in Table \ref{tab:hydrodynamics}) are not merely fitting parameters but have a physical interpretation. The hydrodynamic treatment of shock waves assumes that all the energy lost by the projectile is deposited in a narrow volume around the ion path. The fact that the hot cylinder approximation (a cylinder of 1 nm radius where all the energy is deposited) agrees with the hydrodynamic results shows that 1 nm is a distance short enough to satisfy the assumptions of the model. However, a reduced effective energy loss $S_{\Delta}$ is needed to reproduce the results arising from the use of the radial dose. We can try to identify these $S_{\Delta}$ values with the energies deposited within a certain radius around the ion path, i.e., with a linear energy transfer. Figure \ref{fig:intDr} shows the amount of energy per unit path length deposited within a cylinder of a given radius $\rho$. In the case of 200-keV/u carbon ions the $S_{\Delta}/S = 0.82$ (without $\delta$-electrons) and $S_{\Delta}/S = 0.7$ (with $\delta$-electrons) values are marked by horizontal dotted lines. It can be seen that in both cases the restricted stopping powers correspond to the radius $\rho_{\Delta} \sim 2$ nm. The same is depicted for the 2-MeV/u case where the values $S_{\Delta}/S = 0.66$ (without $\delta$-electrons) and $S_{\Delta}/S = 0.34$ (with $\delta$-electrons) correspond to the radius $\rho_{\Delta} \sim 1$ nm (see Table \ref{tab:hydrodynamics} for the actual values).

From these results we obtain a qualitative rule of thumb for determining the real strength of the shock waves: the wave front features (position and pressure) of the ion-induced shock waves are well characterised by the hydrodynamic model but assuming a restricted stopping power, or LET, $S_{\Delta}$, which corresponds to the energy deposited within the first 1--2 nm from the ion path. Thus, the importance of accounting for an accurate radial dose is very clear: secondary electrons propagate the energy lost by the ion to a certain distance from its path and the more spread out the energy is the more degraded the shock wave is. The vast majority of the secondary electrons ($\sim 75 \%$) are sub-45-eV and they are accurately described by diffusion. The rest of electrons ($\sim 25$ \%) can be regarded as $\delta$-electrons and their contribution to the radial dose can be more easily determined by a spatially restricted LET formula. The more energetic the $\delta$-electrons are (i.e., the more energetic the primary ion is, due to the maximum energy that can be transferred to secondary electrons, $W_{\rm max} = 4 mT/M$) the more spread out the energy is and the more weakened the shock wave will be. This is less important in the Bragg peak region where the contribution from $\delta$-electrons is smaller. However, for larger energy ions $\delta$-electrons have a more important role and the amount of energy transported out of this 1--2 nm radius cylinder is more significant. Therefore shock waves are especially important in the Bragg peak region while their intensity is strongly suppressed out of it due to the lower stopping powers of ions and the larger spread of the deposited energy.



\section{Conclusions}
\label{sec:conclusions}

This work continues a study of ion-induced shock waves on the nanometre scale. These waves, predicted to originate from the paths of high-LET projectiles, are caused by a steep decrease of the radial dose within a few nanometres from the paths. Compared to previous works that only studied the shock waves in the Bragg peak region and approximated the radial dose dependence as a step function, in this work we used the calculated radial doses as initial conditions for the pressure distribution and obtained a more realistic appearance of wave fronts. These radial doses also included the contributions of more energetic $\delta$-electrons which allowed us to extend the simulations to the region out of the Bragg peak.

The radial doses that include the contributions of sub-45-eV electrons that comprise almost 80\% of secondary electrons in the Bragg peak region were calculated before. The transport of those electrons was described by diffusion equations and these provide a good picture of energy distribution in the track $\sim50$~fs after the ion traverse when the track structure is developed. This radial dose distribution is equivalent to the initial pressure distribution, i.e., the structure of the wave front of the shock wave. In this work, these radial doses were augmented with contributions of more energetic $\delta$-electrons. This has been done using a spatially restricted linear energy transfer formula. As a result the large radii tails of the radial dose are in a good agreement with experimental data using tissue-equivalent gas and with Monte Carlo simulation results. The numbers of sub-45-eV and $\delta$-electrons for ions at different energies have been determined by means of the singly differential ionisation inverse mean free path. The addition of the $\delta$-electrons contribution to the radial dose removes an adjustable parameter introduced within the diffusion description, the average energy deposited per inelastic collision, being now fixed to 15 eV.

These radial doses were used to set up the initial conditions for molecular dynamics simulations of the ion-induced shock waves. In this way the validity of the hot cylinder approximation (all the energy uniformly deposited within a 1-nm radius cylinder) used in previous works was evaluated. The main effect of these more realistic simulations is a more gradual wave front. The difference is especially pronounced outside of the Bragg peak region where the effect of $\delta$-electrons becomes significant. The strength of shock waves is influenced by the spread of the wave front. They are not weakened too much in the Bragg peak region where the
assumption of the calculated radial dose compared to a step function energy distribution decreases its intensity by $\sim 30$\% mainly due to the spread of the wave front. However, the radial dose effect is much more dramatic for large ion energies out of the Bragg peak region. For 2-MeV/u carbon ions, the strength of the shock wave (already much less due to the lower stopping power) is weakened by $\sim 65$\%. Therefore, the main conclusion is that the real strength of the shock wave is
determined by the amount of energy deposited within the innermost part of the track, i.e., in the cylinder of radius $\sim$1--2 nm around the ion path.

For the higher ion energies out of the Bragg peak the shock waves are heavily weakened due to the increased influence of $\delta$-electrons on the radial dose. Nonetheless, in the Bragg
peak region, where the contribution from $\delta$-electrons to the radial dose is not very important, most of the energy lost by ions is deposited within this innermost cylinder leading to strong shock waves. All track-structure simulation codes as well as all current biophysical models used
in IBCT, with the exception of the multiscale approach, have ignored the impact that ion-induced shock waves might have on biodamage. While this is permissible for high-energy ions, the effects of shock waves must be taken into account for ions in the Bragg peak region.

\begin{acknowledgement}
\begin{sloppypar}
PdV would like to thank Prof. Isabel Abril and Prof. Rafael Garcia-Molina for providing the stopping power values for ion beams in liquid water and for their advice on the calculation of electron CSDA ranges, as well as them and Dr Maurizio Dapor for interesting discussions on the role of delta electrons on the radial dose. The authors acknowledge financial support from the European Union's FP7-People Program (Marie Curie Actions) within the Initial Training Network No. 608163 ``ARGENT'', Advanced Radiotherapy, Generated by Exploiting Nanoprocesses and Technologies. PdV also acknowledges additional support from the European Regional Development Fund and the Spanish Ministerio de Econom\'{i}a y Competitividad (project No. FIS2014-58849-P). The molecular dynamics simulations were performed using the computer cluster KELVIN at Queen's University Belfast (UK).
\end{sloppypar}
\end{acknowledgement}


%
%

\end{document}